\def\br{{\bf r}}
\def\bR{{\bf R}}
\def\bk{{\bf k}}
\def\bkp{{\bk '}}
\def\bE{{\bf E}}
\def\w0{\omega_0}
\def\k0{k_0}
\def\wk{\omega_k}
\def\wkp{\omega_{k'}}
\def\epkj{\epsilon_{\bk j}}
\def\epkjs{\epsilon_{\bk j}^\star}
\def\epkjp{\epsilon_{\bk 'j'}}
\def\epkjps{\epsilon_{\bk 'j'}^\star}
\def\ekj{\hat{e}_{\bk j}}
\def\ekjs{\hat{e}_{\bk j}^\star}
\def\ekjp{\hat{e}_{\bk 'j'}}

\def\akj{a_{\bk j}}
\def\akjd{a_{\bk j}^\dagger}
\def\akjp{a_{\bk 'j'}}
\def\akjpd{a_{\bk 'j'}^\dagger}
\def\Fln{F_{\ell n}^R}
\def\barR{\bar{R}}
\def\Flm{F_{\ell m}^{\bar{R}}}

\documentclass[preprint,aps,showpacs]{revtex4}

\begin{document}

\title{Dynamical Casimir-Polder energy between an excited and a ground-state atom} 

\author{L. Rizzuto\mbox{${\ }^{*}$}, R. Passante\mbox{${\ }^{**}$}, F. Persico\mbox{${\ }^{*}$}}
\affiliation{
\mbox{${\ }^{*}$} INFM and Dipartimento di Scienze Fisiche ed Astronomiche, 
Universit\'{a} degli Studi di Palermo, Via Archirafi 36, I-90123 Palermo, Italy \\
\mbox{${\ }^{**}$}Istituto di Biofisica - Sezione di Palermo, 
Consiglio Nazionale delle Ricerche, Via Ugo La Malfa 153, I-90146 Palermo, Italy }

\email{roberto.passante@pa.ibf.cnr.it}

\pacs{12.20.Ds }

\begin{abstract}
We consider the Casimir-Polder interaction between two atoms, one in
the ground state and the other in its excited state. The interaction is
time-dependent for this system, because of the dynamical self-dressing
and the spontaneous decay of the excited atom.  We calculate the 
dynamical Casimir-Polder potential between the two atoms using an effective
Hamiltonian approach. The results obtained and their physical
meaning are discussed and compared with previous results based on
a time-independent approach which uses a non-normalizable dressed
state for the excited atom.

\end{abstract}

\maketitle

\section{\label{sec:1}Introduction}

The existence of field fluctuations in the vacuum state is a remarkable 
prediction of quantum field theory.  Vacuum fluctuations produce 
observable effects such as the Casimir force between 
two neutral mirrors or dielectrics 
in the vacuum \cite{BMM01} and the Casimir-Polder force 
between neutral atoms or molecules 
in their ground state \cite{CP48}. 
The Casimir-Polder forces are long-range effects due to the 
interaction of the atoms with the common quantum radiation field.
For intermolecular distances smaller than typical atomic transition
wavelengths from the ground state, they reduce to van der Waals forces;
for larger distances they decrease more rapidly than van der Waals forces
due to retardation effects \cite{CP48,PP87}.
The physical origin of the Casimir-Polder force has been investigated 
in the past in terms of dressed vacuum fluctuations, radiation reaction field 
or vacuum field correlations (for a review see \cite{CPP95}).  
More recent studies have also considered the Casimir-Polder dispersion energy 
between two molecules, one in an excited state and the other in the ground state 
\cite{PT93,PT95,PT95a}. The van der Waals-like interaction between an excited
atom and a dieletric surface has also been considered \cite{FSBC95}.
These calculations are based on fourth-order perturbation theory,
and they are time-independent. In fact, the spontaneous decay 
of the excited atom, as well as its dynamical self-dressing, is not included
in these calculations, the excited atom being treated as it were in a stable
state. The time-independent potential contains two terms: one resulting from
virtual photons exchange, and the other from the resonance due
to the possibility of the emission of a resonant photon \cite{PT95}. 
The term arising from the virtual photons exchange has the same structure 
of the Casimir-Polder potential for ground-state atoms. The resonant term 
is a polynomial in the inverse of the intermolecular separation $R$.
Finally, it has been recently suggested
the possibility of enhancement of van der Waals forces in
nonequilibrium situations \cite{CM03}; this 
indicates that the matter is not entirely settled and explains
our interest in Casimir-Polder forces in dynamical situations. 
The term dynamic in general may refer to two situations, one time-dependent 
and the other frequency dependent, which may also lead to dynamic potentials.
This paper is concerned with the first case, i.e. explicitly time-dependent
situations.

In this paper, we shall adopt a time-dependent approach for the
calculations of the Casimir-Polder potential between a ground-state
and an excited-state atom/molecule. This 
approach, which takes into account both the short time dynamical
dressing and the spontaneous decay of the excited atomic state,
will give a deeper understanding of the physical nature of the
Casimir-Polder force.

As usual,  the interaction energy between the excited  
and the ground-state atom is assumed to stem 
from the response of the latter 
to the field emitted by the former. This idea has recently been
used in the different context of the calculation of the Casimir-Polder force
between partially dressed atoms \cite{PP03}. We use perturbation theory,
and this limits the validity of our results to times shorter than the
lifetime of the excited atom. 
We find that this potential is zero before the ``causality time" $t=R/c$, 
coherently with relativistic causality. 
For $t>R/c$, we find that the interaction energy contains three terms. 
Two of them were already obtained 
in previous time-independent calculations \cite{PT95}. 
The third term is new, and it is  time-dependent; it
describes the time dependence of the force when one atom initially
is in its bare excited state. 
This new term vanishes for times larger than the timescale 
of the dynamical dressing of the excited state, 
which coincides with the so-called Zeno time \cite{POP00}; 
after the Zeno time (but at times shorter than the 
timescale of the spontaneous
decay $\gamma^{-1}$ of the excited atomic state), 
the interaction energy reduces to that obtained 
by time-independent calculations.

The paper is organized as follows. 
In Section \ref{sec:2} we describe our effective Hamiltonian
approach, and in Section \ref{sec:3}
we obtain the complete Casimir-Polder potential 
between the excited and the ground-state atom, inclusive of
the old (time-independent \cite{PT95}) and of the new
(time-dependent) terms.

\section{\label{sec:2}The effective Hamiltonian}

We consider two atoms A and B interacting with the 
electromagnetic radiation field in the Coulomb gauge;  ${\bf r}_{A}$ 
and ${\bf r}_{B}$ are their position.
Atom A is approximated as a two-level systems. Its interaction
with the radiation field, in the multipolar coupling scheme and 
within dipole approximation, is described by the following Hamiltonian 
\cite{CPP95}

\begin{eqnarray}
H &=& \hbar \w0 S_{z}^{A} +
\sum_{\bk j} \hbar \wk \akjd \akj + \nonumber
\\
&+& 
\sum_{\bk j} \left( \epkj S_{+}^{A} - \epkjs S_{-}^{A} \right)
\left( \akj e^{i{\bf k} \cdot r_{A}} - \akjd e^{-i{\bf k} \cdot r_{A}} 
\right) 
\label{eq:1}
\end{eqnarray}         
where $\w0 =c\k0$ is the transition frequency of the atom and
$S_{z}$, $S_{+}$ and $S_{-}$ are the pseudospin atomic 
operators.
The coupling constant $\epkj$, in the multipolar
coupling scheme is given by  

\begin{equation}
\epkj = i \left( \frac{2\pi\hbar c k}{V} \right)^{1/2} 
\mbox{\boldmath $e$}_{{\bf k}j}\cdot \mbox{\boldmath $\mu$}^A
\label{eq:2}
\end{equation}
where $\mbox{\boldmath $\mu$}^A$ is the transition dipole
moment of atom A and $\ekj$ are the polarization unit vectors.

The use of the multipolar form of the interaction Hamiltonian
is very convenient in our calculation. In fact, 
in this coupling scheme the momentum
conjugate to the vector potential is the transverse displacement
field which, outside the atoms, coincides with the total 
electric field \cite{PZ59} (transverse plus longitudinal).
In this way, we directly obtain the total field generated
by one atom, inclusive of the longitudinal components.

We assume that at $t=0$ the atom A is in its bare excited state,
while the atom B is in the ground state. The two atoms are in general different,
and we consider a factorized state as initial state.
We are interested in the dynamical Casimir-Polder potential between 
these two atoms.  Our calculation proceeds in two steps. First, we obtain
the electromagnetic field emitted by the initially excited 
atom A, and then we evaluate the interaction energy of
the ground state atom B with this field.
We have already used a similar procedure to obtain the
Casimir-Polder potential between ground-state atoms and
shown its relation with the spatial correlations of vacuum
fluctuations \cite{PPR03}.

The interaction energy of the ground-state atom B with the field
emitted by the excited atom A can be conventiently obtained
by an effective interaction, which is quadratic in the field operators.
The two atoms are in general different.
This quadratic coupling can be obtained by a unitary transformation
from the multipolar Hamiltonian, and it is given by \cite{PPT98,PT83}

\begin{eqnarray}
H_{eff} &=& -\frac 12 \sum_{\bk j} \alpha^B(k)  \langle \bE_{\bk j}(\br_B,t)
\cdot \bE (\br_B,t) \rangle
\nonumber \\
&=& -\frac 12 \sum_{\bk j} \sum_{\bk 'j'} \alpha^B(k) \langle \bE_{\bk j} (\br_B,t) 
\cdot \bE_{\bk 'j'}(\br_B,t) \rangle
\label{eq:3}
\end{eqnarray}
where the average in (\ref{eq:3}) has to be taken on the
initial state of the system (atom A excited and 
the field in the vacuum state), $\alpha^B(k)$ is the ground-state 
dynamic polarizability of the atom B and

\begin{equation}
\bE (\br_B,t) = \sum_{\bk j} \bE_{\bk j}(\br_B,t) =
i \sum_{\bk j} \sqrt{\frac{2\pi \hbar \wk}V} \left( \ekj \akj (t)
e^{i\bk \cdot \br_B} -  \ekjs \akjd (t) e^{-i\bk \cdot \br_B} \right)
\label{eq:4}
\end{equation}
is the field operator evaluated at the position of atom B, 
$ \bE_{\bk j}(\br_B,t)$ being its $(\bk j)$ component, which
includes a contribution coming from the presence
of atom A. In this way, we 
obtain the Casimir-Polder potential between the atoms A and B from
the response of atom B to the field emitted by atom A.
We stress that 
the field operator $\bE$ in (\ref{eq:4}) is the transverse displacement
field operator (that is, the momentum conjugate to the vector potential)
which, outside the atoms, coincides with the total electric field
operator \cite{PZ59}: longitudinal field contributions are already
included in (\ref{eq:4}).

\section{\label{sec:3}The dynamical Casimir-Polder potential}

The first step to obtain the time-dependent Casimir-Polder potential, as 
outlined above, is to evaluate the average value of the
operator $\bE_{\bk j}(\br_B,t) \cdot \bE_{\bkp j'}(\br_B,t)$ on the initial state,
that is the state with atom A excited and the field in
the vacuum state. We obtain this quantity by solving at the second order in
the coupling constant the Heisenberg
equations of motion for the field operators and
using the Hamiltonian (\ref{eq:1}), and then taking the average value on
the state at $t=0$; the calculation is sketched out in 
Appendix \ref{AppendixA}. 
Our procedure follows closely that by Power and Thirunamachandran
\cite{PT83} for a multilevel atom, with the difference that we have
specialized to a two-level case and that we have dealt explicitly
with tha case $t<R/c$.
Substitution of (\ref{eq:A6}) into
(\ref{eq:3}) yields the following expression for the average value of 
$H_{eff}$, which gives the Casimir-Polder potential between
the two atoms,

\begin{eqnarray}
\Delta E_{AB} &=& -\frac 12 \sum_{\bk j \bkp j'} \alpha^B(k)
\langle \uparrow_A \{ 0_{\bk j} \} 
\mid \bE_{\bk j}(\br_B,t) \cdot \bE_{\bkp j'}(\br_B,t)
\mid \uparrow_A \{ 0_{\bk j} \} \rangle
\nonumber \\
&=& \frac 12 \left( \frac {2\pi c}V \right)^2 
\sum_{\bk j \bkp j'}
\left( \ekj \cdot \ekjp \right)
\left( \ekj \cdot \mbox{\boldmath $\mu$}^A \right)
\left( \ekjp \cdot \mbox{\boldmath $\mu$}^A \right)
\nonumber \\
&\times& \Bigg\{ 
\alpha^B(k) \left( F_t(\w0 +\wk ) e^{i(\bk \cdot \bR - \wk t)}
- F_t(\w0 -\wk ) e^{-i(\bk \cdot \bR - \wk t)} \right)
\nonumber \\
&\times& \left( F_t^\star (\w0 -\wkp ) e^{i(\bkp \cdot \bR - \wkp t)}
- F_t^\star (\w0 +\wkp ) e^{-i(\bkp \cdot \bR - \wkp t)} \right)
\nonumber \\
&+& i\alpha^B(k)e^{i(\bk \cdot \bR -\wk t)}
\Bigg[  \frac 1{\w0 -\wk} \bigg( e^{i(\bkp \cdot \bR -\wkp t)}
\left( F_t(\wk +\wkp ) - F_t(\w0 +\wkp ) \right)
\nonumber \\
&-& e^{-i(\bkp \cdot \bR -\wkp t)}
\left( F_t(\wk -\wkp ) - F_t (\w0 -\wkp ) \right) \bigg)
\nonumber \\
&+&   \frac 1{\w0 +\wk} \bigg( e^{i(\bkp \cdot \bR -\wkp t)}
\left( F_t(\wk +\wkp ) - F_t^\star (\w0 -\wkp ) \right)
\nonumber \\
&-& e^{-i(\bkp \cdot \bR -\wkp t)}
\left( F_t(\wk -\wkp ) - F_t^\star (\w0 +\wkp ) \right) \bigg)
 \Bigg]
\nonumber \\
&-& i\alpha^(k')e^{-i(\bk \cdot \bR -\wk t)}
\Bigg[  \frac 1{\w0 -\wk} \bigg( e^{-i(\bkp \cdot \bR -\wkp t)}
\left( F_t^\star (\wk +\wkp ) - F_t^\star (\w0 +\wkp ) \right)
\nonumber \\
&-& e^{i(\bkp \cdot \bR -\wkp t)}
\left( F_t^\star (\wk -\wkp ) - F_t^\star (\w0 -\wkp ) \right) \bigg)
\nonumber \\
&-& \frac 1{\w0 +\wk} \bigg( e^{-i(\bkp \cdot \bR -\wkp t)}
\left( F_t^\star (\wk +\wkp ) - F_t (\w0 -\wkp ) \right)
\nonumber \\
&-& e^{i(\bkp \cdot \bR -\wkp t)}
\left( F_t^\star (\wk -\wkp ) - F_t (\w0 +\wkp ) \right) \bigg)
 \Bigg] \Bigg\}
\label{eq:5}
\end{eqnarray}
where the complex function $F_t(x)$ is defined in (\ref{eq:A5}).

We first perform integrations/summations over $\bk 'j'$ in the
continuum limit, obtaining 

\begin{eqnarray}
& & \Delta E(A,B) = \frac \pi V \mbox{\boldmath $\mu$}^A_{m}
\mbox{\boldmath $\mu$}^A_{n}
\sum_{\bk j} \left( \ekj \right)_\ell \left( \ekj \right)_m
\nonumber \\
&\times& \Bigg\{ \alpha^B(k) ic  \left( F_t(\w0 -\wk ) e^{-i(\bk \cdot \bR -ckt)}
- F_t(\w0 +\wk ) e^{i(\bk \cdot \bR -ckt)} \right)
e^{-i\k0 ct} \Fln \frac {e^{i\k0 R}}R + \alpha^B(k) e^{i\bk \cdot \bR}
\nonumber \\
&\times&  \Bigg[ \frac 1{\k0 -k} 
\left( \Fln \frac {e^{-ikr}}R - e^{i(\k0-k)ct} \Fln \frac {e^{-i\k0 r}}R \right) 
+ \frac 1{\k0 +k} 
\left( \Fln \frac {e^{-ikr}}R - e^{-i(\k0+k)ct} \Fln \frac {e^{i\k0 r}}R \right) \Bigg]
\nonumber \\
&+& e^{-i\bk \cdot \bR}\Bigg[ \frac 1{\k0 -k} 
\left( \alpha^B(k) \Fln \frac {e^{ikr}}R - \alpha^B(\k0 ) e^{-i(\k0-k)ct} 
\Fln \frac {e^{i\k0 r}}R \right)
\nonumber \\ 
&+& \frac 1{\k0 +k} \left( \alpha^B(k) \Fln \frac {e^{ikr}}R 
- \alpha^B(\k0 ) e^{i(\k0 +k)ct} \Fln \frac {e^{-i\k0 r}}R \right) \Bigg]
\Bigg\} \Theta (ct-R)
\label{eq:6}
\end{eqnarray} 
where we have defined the differential operator acting on the 
variable $\bR$
\begin{equation}
\Fln = \left( -\delta_{n\ell} \nabla^2 + \nabla_\ell \nabla_n \right)
\label{eq:7}
\end{equation}

The presence of the $\Theta$ function in (\ref{eq:6}) ensures relativistic
causality in the propagation of the field generated by atom A and
consequently in the interaction between the two atoms. The $\Theta$
function results from integrals over $k$ of the following kind
\begin{equation}
P\int_{-\infty}^\infty dk \frac {e^{ikx}}{k+\k0} \alpha (k)
=i\pi (2\Theta (x)-1) e^{-i\k0 x} \alpha (\k0 )
\label{eq:7b}
\end{equation}

After lengthy calculations which include integration over $\bk j$
of part of the terms containing $1/(\k0 -k)$, 
equation (\ref{eq:6}) can be expressed
in the more compact form

\begin{eqnarray}
& & \Delta E(A,B) = 
\frac{2\pi}V \mbox{\boldmath $\mu$}^A_{m}
\mbox{\boldmath $\mu$}^A_{n}
\sum_{\bk j} \left( \ekj \right)_\ell  \left( \ekj \right)_m
\frac k{\k0 -k}
\nonumber \\
&\times&  \Re \left\{ e^{i\bk \cdot \bR} 
\left( 2\alpha^B(k) \Fln \frac {e^{-ikr}}R 
-\left( \alpha^B(k) + \alpha^B(\k0 ) \right)
e^{i(\k0 -k)ct} \Fln  \frac {e^{-i\k0 r}}R \right) \right\}
\Theta (ct-R)
\label{eq:7a}
\end{eqnarray}

After summation over $(\bk j)$ in the continuum limit
and some algebraic manipulations where the analytical
properties of the dynamical polarizability $\alpha^B(k)$
are used, we finally get

\begin{eqnarray}
\Delta E(A,B) = &\Bigg\{& -\mbox{\boldmath $\mu$}^A_{m}
\mbox{\boldmath $\mu$}^A_{n}
\alpha^B(\k0 ) \Fln \frac 1R \Flm \frac 1\barR
\cos \k0 (R -\barR )  
\nonumber \\
&+& \frac {\hbar c}{2\pi}  \Fln \frac 1R \Flm \frac 1\barR
\int_0^\infty du e^{-u(R+\barR )} \alpha^A_{mn}(iu) \alpha^B(iu)
+ \frac 1\pi \mbox{\boldmath $\mu$}^A_{m}
\mbox{\boldmath $\mu$}^A_{n} \Fln \frac 1R \Flm \frac 1\barR
\nonumber \\
&\times& \Bigg[ \cos \k0 (ct-R) \int_0^\infty du
\left( \alpha^B(iu) + \alpha^B(iu_0) \right) e^{-uct}
\frac {2\k0 \sinh u \barR}{\k0^2 + u^2} + \sin \k0 (ct-R)
\nonumber \\
&\times&  \int_0^\infty du
\left( \alpha^B(iu) + \alpha^B(iu_0) \right) e^{-uct}
\frac {2u \sinh u \barR}{\k0^2 + u^2} \Bigg]
\Bigg\}_{R=\barR} \Theta (ct-R)
\label{eq:8}
\end{eqnarray}
where the variable $\barR$, which is put equal to $R$ after
the action of the differential operator $\Flm$, has been
conveniently introduced in order to distinguish the variables on
which the operators $\Fln$ and $\Flm$ operate.
$\alpha_A(iu)$ is the dynamical polarizability of the excited state of 
the atom A, extended to imaginary frequencies $iu$,
\begin{equation}
\alpha^A_{mn}(iu) = \frac {2\k0 \mbox{\boldmath $\mu$}^A_{m}
\mbox{\boldmath $\mu$}^A_{n}}
{\hbar c (\k0^2 +u^2)}
\label{eq:9}
\end{equation}
For a two-level system, the dynamical polarizability of the excited
state coincides with that of the ground state except for a change of
its sign. Experimental observability of time dependences of the form
implied by expression (\ref{eq:8}) has been discussed in \cite{CPPP90}.

We notice from equation (\ref{eq:8}) that the first two terms inside the curly
bracket are time-independent, whereas the third term depends on time. 
This time-dependent term contains, inside the $u$-integrals, an exponential
factor decreasing with time. For a given $R$, this term
rapidly vanishes to zero with a time-scale of the order of $\k0^{-1}/c = \w0^{-1}$.
This means that for this given $R<ct$ 
after a transient in which there is a time-dependent
Casimir-Polder interaction, then the interatomic interaction stabilizes to
\begin{eqnarray}
\Delta E(A,B) = &\Bigg\{& -\mbox{\boldmath $\mu$}^A_{m}
\mbox{\boldmath $\mu$}^A_{n}
\alpha_B(\k0 ) \Fln \frac 1R \Flm \frac 1\barR
\cos \k0 (R -\barR )  
\nonumber \\
&+& \frac {\hbar c}{2\pi}  \Fln \frac 1R \Flm \frac 1\barR
\int_0^\infty du e^{-u(R+\barR )} \alpha^A_{mn}(iu) \alpha^B(iu)
\Bigg\}_{R=\barR}
\label{eq:10}
\end{eqnarray}
which is time-independent. We note that
the timescale $\w0^{-1}$ of the dynamical Casimir-Polder potential
is the same of the nonexponential early stage of the spontaneous decay
of the excited atom (Zeno time). 
Details of the time-dependent term in (\ref{eq:8}) may dipend from
the choice of the initial state at $t=0$, in our case a bare excited state.
Other possible choices, for example a partially dressed state, might yield a
different expression of this term, but we expect that the general
properties of the time-dependent energy should not change.
The time-dependent energy in (\ref{eq:8}) yields a time-dependent
force between the two atoms, that in principle is observable.
During this stage of the decay,
the self dressing of the atom occurs \cite{CPPP95}. This indicates
that the time-dependent part of the potential is related to the interaction
of atom B with the dynamical photon cloud of atom A which is generated
during its self-dressing.
Our result (\ref{eq:8}) is valid only up to
times of the order of $\gamma^{-1}$ or smaller, where $\gamma$ is the decay
rate of the excited state, because of the limitation of the perturbation
theory we have used. However, for atomic systems the time interval between
$\k0^{-1}/c$ and $\gamma^{-1}$ is typically quite a long interval. 
Equation (\ref{eq:10}) coincides with the result obtained by
Power and Thirunamachandran using a time-independent approach
based on a non-normalizable dressed excited state for atom A
\cite{PT95,PT95a}. This part of the potential has two components:
one has the same form of the potential for ground state atoms, and
the other is spatially oscillating and it is related to the fact that the
excited atom can emit a resonant photon.

\section{\label{sec:4}Conclusions}

We have considered the Casimir-Polder intermolecular interaction between
two atoms, one in its ground state and the other excited. The latter
is assumed to be at $t=0$ in its bare excited state. We have used an effective
Hamiltonian approach, and the interaction energy between the two atoms
stem from the interaction of the ground state atom (through its dynamical
polarizability) with the field generated by the excited atom. The interaction
energy yielding the Casimir-Polder potential is time-dependent because
of the dynamical self-dressing processes of the excited atom; there is also
a contribution to the potential from the resonance related to the possibility
of emission of a resonant photon by the excited atom. We find that for times
$t \gg \w0^{-1}$, that is for times larger than the inverse of the transtion
frequency of the excited atom, and for $t>R/c$,
the Casimir-Polder interaction becomes
time-independent. In this limit its expression coincides with that already
obtained by Power and Thirunamachandran using a time-independent
approach and based on a non-normalizable dressed state for the excited
atom. We argue that the time-dependent part of the potential that we obtain
is due the virtual photons which are emitted by the excited  atom in the
very early stages of its decay.

\begin{acknowledgments}
This work was supported by the European Commission under contract No. HPHA-CT-2001-40002
and in part by the bilateral Italian-Japanese project 15C1 on Quantum
Information and Computation of the Italian Ministry for Foreign Affairs. 
Partial support by Ministero dell'Universit\'{a} e della Ricerca Scientifica 
e Tecnologica and by Comitato Regionale di Ricerche Nucleari e di Struttura della Materia
is also acknowledged.
\end{acknowledgments}

\appendix
\section{\label{AppendixA}Iterative solution of the Heisenberg equations}

In this Appendix we outline the iterative solution of the Heisenberg
equations describing the interaction of atom A with the radiation field,
using the Hamiltonian (\ref{eq:1}) for the part pertaining to atom A.
The Heisenberg equations for the field and atomic operators are

\begin{equation}
\dot{a}_{\bk j}(t) = -i\wk \akj (t) + \frac i\hbar \left( \epkj S^A_+(t)
-\epkjs S^A_-(t) \right) e^{-i\bk \cdot \br_A}
\label{eq:A1} 
\end{equation}

\begin{equation}
\dot{S}^A_+(t) = i\w0 S^A_+(t) +\frac {2i}\hbar S^A_z(t) \sum_{\bk j} \epkjs
\left( \akj (t) e^{i\bk \cdot \br_A} - \akjd (t) e^{-i\bk \cdot \br_A} \right)
\label{eq:A2} 
\end{equation}

The iterative solution of these equations and their Hermitian
conjugates yields the perturbative expansion of the
field operators

\begin{equation}
\akj (t) = \akj^{(0)}(t) +  \akj^{(1)}(t) + \akj^{(2)}(t) + \ldots
\label{eq:A3} 
\end{equation}
where
\begin{eqnarray}
\akj^{(0)}(t) &=& \akj (0) e^{-i\wk t}
\label{eq:A4a} \\
\akj^{(1)}(t) &=& \frac i\hbar e^{-i\wk t} \left( \epkj S^A_+(0) 
F_t^\star (\w0 +\wk ) - \epkjs S^A_-(0) F_t^\star (\w0 -\wk ) \right)  e^{-i\bk \cdot \br_A}
\label{eq:A4b} \\
\akj^{(2)}(t) &=& -\frac {2i}{\hbar^2} S^A_z(0) e^{-i\wk t} e^{-i\bk \cdot \br_A}
\sum_{\bk ' j'} \Bigg\{  \akjp (0) e^{i\bk ' \cdot \br_A}
\nonumber \\
&\times& \left( \epkj \epkjps \frac {F_t(\w0 -\wkp ) - F_t(\w0 +\wk )}{\w0 +\wkp}
+ \epkjs \epkjp \frac {F_t(\wk -\wkp ) - F_t^\star (\w0 -\wk )}{\w0 -\wkp} \right) 
\nonumber \\
&-&  \akjpd (0) e^{-i\bk ' \cdot \br_A} \left( \epkj \epkjps
\frac {F_t(\w0 +\wkp ) - F_t(\w0 +\wk )}{\w0 -\wkp} \right.
\nonumber \\
&+& \left. \left.  \epkjs \epkjp \frac {F_t(\wk +\wkp ) - F_t^\star (\w0 -\wk )}{\w0 +\wkp}
\right) \right\}
\label{eq:A4c}
\end{eqnarray}
where we have defined the function
\begin{equation}
F_t(x) = \int_0^t  \! dt' e^{ixt'}
\label{eq:A5}
\end{equation}

Using (\ref{eq:A4a}, \ref{eq:A4b}, \ref{eq:A4c}), we obtain the following
expression of the average value of the field operators
present in equation (\ref{eq:3}) on the initial state of 
the system (atom A + field) $\mid \uparrow_A \{ 0_{\bk j} \} \rangle$ 

\begin{eqnarray}
&\ & \langle \bE_{\bk j} (\br_B,t) \cdot \bE_{\bk j} (\br_B,t) \rangle
= - \left( \frac {2\pi c}V \right)^2 \left( \ekj \cdot \ekjp \right)
\left( \ekj \cdot \mbox{\boldmath $\mu$}^A \right)
\left( \ekjp \cdot \mbox{\boldmath $\mu$}^A \right) kk'
\nonumber \\
&\times& \Bigg\{ \left( F_t(\w0 +\wk ) e^{i(\bk \cdot \bR - \wk t)}
- F_t(\w0 -\wk ) e^{-i(\bk \cdot \bR - \wk t)} \right)
\nonumber \\
&\times& \left( F_t^\star (\w0 -\wkp ) e^{i(\bk ' \cdot \bR - \wkp t)}
- F_t^\star (\w0 +\wkp ) e^{-i(\bk '\cdot \bR - \wkp t)} \right)
\nonumber \\
&+& \bigg[ -\left( \frac{F_t(\wk +\wkp ) - F_t(\w0 +\wkp )}{i(\w0 -\wk )}
+ \frac{F_t(\wk +\wkp ) - F_t^\star (\w0 -\wkp )}{i(\w0 +\wk )} \right)
e^{i(\bk +\bk ') \cdot \bR -i(\wk +\wkp ) t} 
\nonumber \\
&+& \left( \frac{F_t(\wk -\wkp ) - F_t(\w0 -\wkp )}{i(\w0 -\wk )}
+ \frac{F_t(\wk -\wkp ) - F_t^\star (\w0 +\wkp )}{i(\w0 +\wk )} \right)
e^{i(\bk -\bk ') \cdot \bR -i(\wk - \wkp ) t} 
\nonumber \\
&+& c.c. ( \bk \leftrightarrow \bk ' ) \bigg] \Bigg\}
\label{eq:A6}
\end{eqnarray}
(the last term indicates the complex conjugate of the terms inside 
the square bracket after exchange between $\bk$ and $\bk '$).  
$\bR = \br_B - \br_A$ is the interatomic separation.

\end{document}